# Autonomic management of multiple non-functional concerns in behavioural skeletons

Marco Aldinucci°, Marco Danelutto⋄ & Peter Kilpatrick⋆

°Dept. Computer Science – Univ. Torino
⋄Dept. Computer Science – Univ. Pisa
⋆Dept. Computer Science – Queen's Univ. Belfast

# Autonomic management of multiple non-functional concerns in behavioural skeletons

Marco Aldinucci°, Marco Danelutto⋄ & Peter Kilpatrick⋆

°Dept. Computer Science – Univ. Torino
⋄Dept. Computer Science – Univ. Pisa
⋆Dept. Computer Science – Queen's Univ. Belfast

**Abstract**

We introduce and address the problem of concurrent autonomic management of different non-functional concerns in parallel applications build as a hierarchical composition of behavioural skeletons. We first define the problems arising when multiple concerns are dealt with by independent managers, then we propose a methodology supporting coordinated management, and finally we discuss how autonomic management of multiple concerns may be implemented in a typical use case. The paper concludes with an outline of the challenges involved in realizing the proposed methodology on distributed target architectures such as clusters and grids. Being based on the behavioural skeleton concept proposed in the CoreGRID GCM, it is anticipated that the methodology will be readily integrated into the current reference implementation of GCM based on Java ProActive and running on top of major grid middleware systems.

⁰This research has been partially carried out under the FP6 Network of Excellence CoreGRID and the FP6 GridCOMP project funded by the European Commission (Contract IST-2002-004265 and FP6-034442) and it is currently being carried out under the CoreGRID ERCIM Working Group ("Advanced programming models" topic).

# 1  Introduction

Efficient implementation of parallel/distributed applications requires solving several problems related to the handling of different non-functional concerns. A *non-functional concern* is a concern not related to *what* is computed by the application, but rather to *how* the results of the application are computed [1] [10]. Typical examples of non-functional concerns include performance tuning, fault tolerance, security and power efficiency.

Effective handling of non-functional concerns requires substantial programming effort. Moreover, the kind of knowledge required to handle non-functional concerns differs completely from the (core) knowledge of *application* programmers. Typically, concrete target architecture knowledge is required, together with deep knowledge of the particular techniques used to handle the non-functional concerns.

In [2] we discussed a framework based on the concept of *behavioural skeleton*, aimed at supporting the programming of parallel/distributed applications. A behavioural skeleton is a co-designed and optimized implementation of a parallel algorithmic skeleton modelling a well-known parallelism exploitation pattern, *together* with an autonomic manager taking care of one of the non-functional concerns related to the execution of that algorithmic skeleton. The complete behavioural skeleton framework has been experimented within the GCM context[2] [8]. Simple managers, taking care of some non-functional concern in a single behavioural skeleton have been designed and implemented [2], as well as hierarchies of autonomic managers, each taking care of a single non-functional concern relative to a single skeleton in a hierarchy of skeletons [3, 5]. In both cases, experimental results demonstrated the feasibility of the behavioural skeleton approach and the efficiency of the GCM implementation of behavioural skeleton in its application to real use cases.

However, the autonomic management of *multiple* non-functional concerns has not yet been considered in this framework, although it is clear that it would be a very useful and powerful tool to tackle non-functional concerns[3].

When dealing with autonomic management of multiple non-functional concerns, several distinct issues arise, in addition to those for a single non-functional concern. In particular, *coordination* of the autonomic managers taking care of the different concerns is needed, to avoid conflicting decisions being taken that eventually impair the whole autonomic management framework. This coordination represents a significant challenge.

---

[1] Wikipedia gives a slightly different definition of this: *In general, functional requirements define what a system is supposed to **do** whereas non-functional requirements define how a system is supposed to **be**.* (http://en.wikipedia.org/wiki/Non-functional_requirement)

[2] http://gridcomp.ercim.org/content/view/26/34/

[3] quoting [13]: *In larger systems with a hierarchical architecture, managers must be able to interact with each other.*

In this paper we consider autonomic management of several different non-functional concerns in a distributed system. We address the problem in a structured programming framework (Sec. 2), we consider the issues related to coordination of autonomic managers each dealing with a different concern (Sec. 3), and we discuss the methodology proposed in Sec. 3.1 applied to a typical use case (Sec. 4). Finally, related work and conclusions and future work sections conclude the paper.

## 2  Parallel framework

We assume here that parallel applications are programmed according to *structured parallel programming principles* [7]. In particular, we assume a parallel application is build of the composition of *behavioural skeletons* [1] and sequential portions of code modelling pure functions. A behavioural skeleton (BS) models a well-know parallelism exploitation pattern. We assume here the existence of a set of BS including: *pipeline* modelling computations in stages processing streams of tasks; *task farm* modelling embarrassingly parallel computations processing streams of tasks; *data parallel* modelling different kinds of data parallel patterns (embarrassingly parallel, with stencil, with shared read only data structures, etc.); and *sequential* wrapping pure functional sequential code in such a way that it can be used within other BS. Each BS implements a known parallelism exploitation pattern *and* an autonomic manager taking care of some non-functional concern. A parallel application is thus build out of a composition of BS. The user provides the sequential portions of code wrapped in the sequential BS, the input data *and* a QoS contract. The BS run time executes the application in such a way that the (hierarchy of) application manager(s) takes care of ensuring the QoS contract provided by the user.

As an example, in [3] we discuss an application which is a pipeline whose first and third stages are sequential, whose second stage is parallel (a task farm with sequential workers) and whose autonomic manager deals with performance tuning. The structure of the resulting application is depicted in Fig. 1.

Restriction of the parallelism patterns the programmer can exploit by the use of behavioural skeletons makes it possible to achieve better performance and efficiency while implementing the application, and allows effective autonomic management to be programmed in the autonomic managers[4] while preserving the possibility to model all (most) of the commonly used patterns in parallel and distributed computing.

---

[4]as the parallel structure of the application is completely known

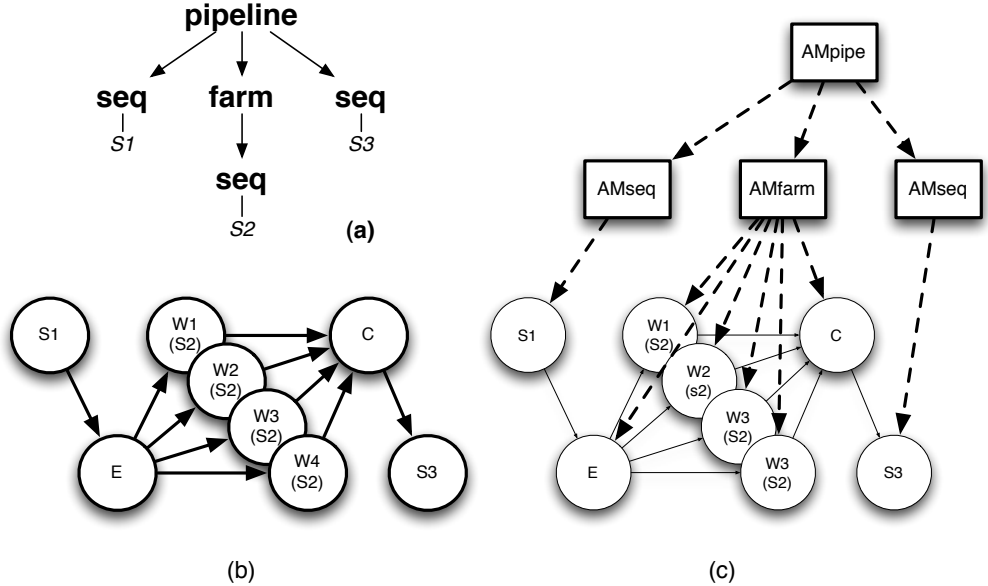

Figure 1: Sample parallel application with behavioural skeletons: logic view (a), process view (sample, (b)) and autonomic manager view (c)

# 3 Autonomic management of multiple concerns in structured parallel computations

When dealing with multiple non-functional concerns, we have to consider that, in the most general cases, distinct autonomic management strategies may exist for each of the non-functional concerns under consideration. More precisely, we may assume that a collection of (possibly hierarchical) autonomic managers exist $\mathcal{AM}_1, \ldots, \mathcal{AM}_m$ that can independently and autonomously take care of non-functional concerns $\mathcal{C}_1, \ldots, \mathcal{C}_m$. For example, the managers AMpipe, AMseq (two instances) and AMfarm of Fig. 1 constitute a single, hierarchically structured collection of autonomic managers. If more concerns are to be considered, we will assume more managers will be associated with the single behavioural skeleton. Fig. 2 shows how these managers will be organized when two non-functional concerns are involved: $\mathcal{C}_P$ (performance tuning) and $\mathcal{C}_S$ (security).

Hereafter, we will use the term $\mathcal{AM}_i$ to refer to the top level manager of a hierarchy of managers handling non-functional concern $\mathcal{C}_i$, if not otherwise specified.

Our approach to handling multiple non-functional concerns is based on a five-pronged attack: identifying an overall strategy for coordinating the managers' activities; finding a common currency by which managers may interact; finding means of reaching consensus on decisions; determining how the management activity can be initialized; and devising a means to implement autonomic

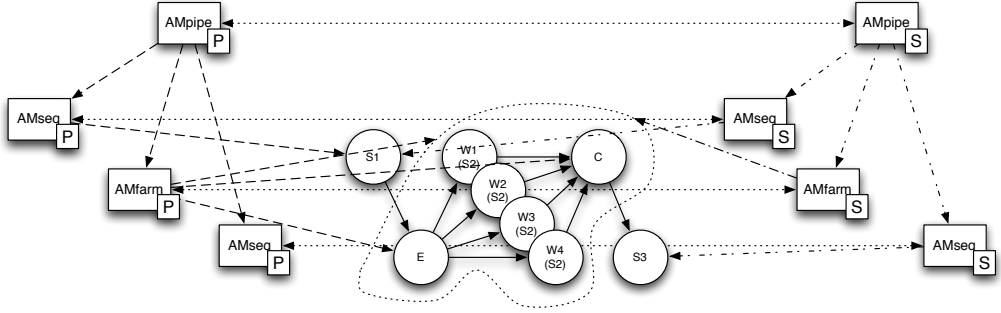

Figure 2: Multiple manager hierarchies (S=security managers, P=performance tuning managers) in behavioural skeletons

management. We now consider each of these in turn.

## 3.1 Centralized vs. distributed autonomic management of multiple concerns

When considering autonomic management of *multiple* non-functional concerns, we must identify a general strategy to coordinate the autonomic management activities performed by the different managers (or manager hierarchies). In general, it may be the case that manager $\mathcal{AM}_i$ takes a decision affecting the global application that is in contrast with the strategies of manager $\mathcal{AM}_j$. For example, $\mathcal{AM}_P$ (a manager taking care of ensuring performance contracts) may clearly take decisions that are in contrast with the policies ensured by $\mathcal{AM}_W$ (the manager taking care of ensuring power management contracts).

To resolve these conflicts, a means must exist by which managers may reach mutually acceptable positions. Two strategies can be identified for this purpose:

SM a *Super Manager* $\mathcal{AM}_0$ can be introduced, positioned hierarchically above managers $\mathcal{AM}_1$ to $\mathcal{AM}_m$, coordinating the decisions taken locally by these autonomic managers and relating to different, possibly interfering concerns; or

CM the managers $\mathcal{AM}_1$ to $\mathcal{AM}_m$ can be modified in such a way that *before* actuating any decision taken, they reach agreement with the others.

Both solutions share a common concept, which is the idea of building a *consensus* on the decisions taken. In the former case (SM) the consensus has to be sought by $\mathcal{AM}_0$, upon communication from one of the $\mathcal{AM}_i$ of a proposed decision. Upon consensus, $\mathcal{AM}_0$ may give the green light to $\mathcal{AM}_i$ in such a way that the decision is actuated. If consensus is not reached, eventually $\mathcal{AM}_0$ will communicate to the $\mathcal{AM}_i$ that the decision is to be aborted. In the latter case (CM), the $\mathcal{AM}_i$ that proposes to take a decision should contact all the other managers and behave as the SM in the former case to build a consensus on this decision. So, the two strategies considered differ only in the way they will build the manager network, but thereafter most of the coordination algorithms and strategies should be the same, or very similar.

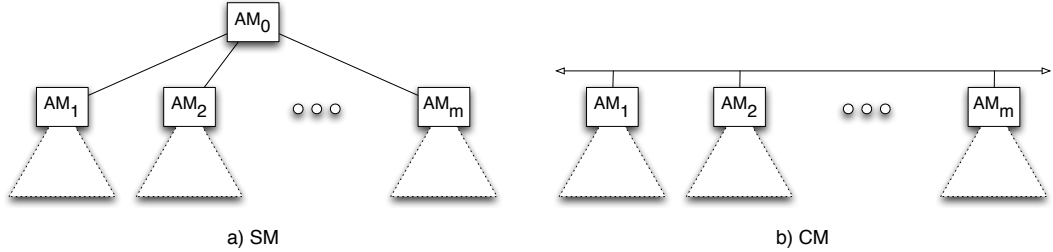

Figure 3: Alternative strategies to implement global multi-concern management

As a matter of fact, in solution CM the coordination among managers may happen at any level of the autonomic manager hierarchy. Fig. 3 exemplifies the two extreme cases. Fig. 2 shows how managers dealing with different concerns within the same behavioural skeleton can be naturally paired in such a way they can coordinate locally taken decisions.

## 3.2 Shared knowledge among different autonomic managers

The second area to be addressed when reasoning about multi-concern management is the common knowledge necessary across different concern managers to make possible agreement on global application management. Different manager hierarchies should agree on a common view of the parallel/distributed application at hand in order to be able to share decisions and, where appropriate, obtain consensus on local decisions before actuating them.

The main common concept across the different managers is the *application graph* whose nodes represent the parallel/distributed activities and whose arcs represent communications/synchronizations among these activities. Each node and arc can be labelled with suitable *metadata*. For example, the node metadata could represent *mapping* information (which processing element(s) host the parallel activity, what are its features in terms of CPU, memory, disk, network bandwidth, etc.); the arc metadata may represent features of the corresponding communication channel (kind of protocol used, bandwidth and latency, whether it can be regarded as a secure channel or not, etc.).

We do not address here general parallel/distributed applications. Rather, we target only those applications build by composing behavioural skeletons[5]. Therefore the application graph we will deal with is the graph representing a *well*

---

[5]actually, of the algorithmic skeletons modelled by the behavioural skeletons

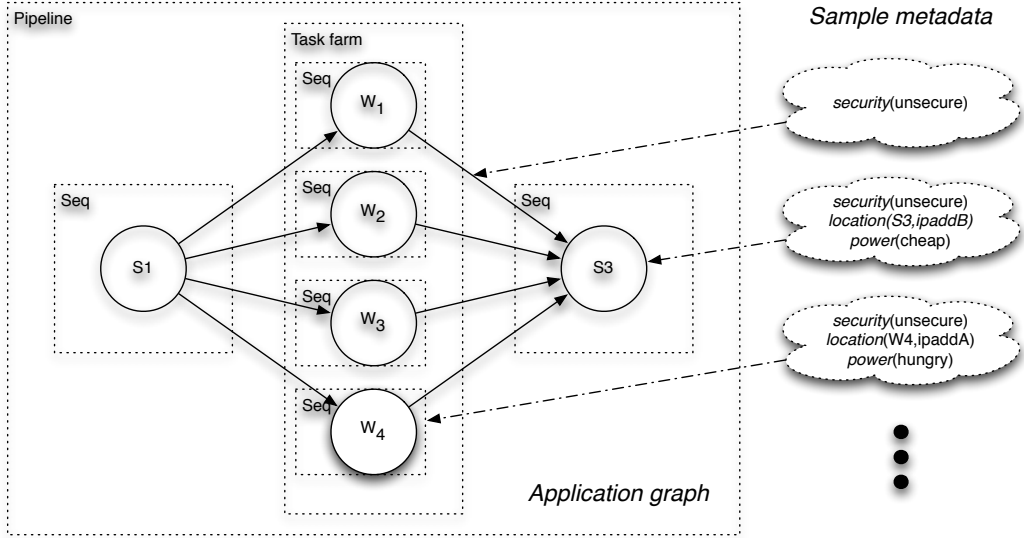

Figure 4: Sample application graph

*formed composition* of parallel/distributed patterns modelled by the behavioural skeleton library at hand. Fig. 4 shows the application graph (with sample associated metadata) corresponding to our sample application: a three stage pipeline with parallel second stage (task farm with 4 workers).

The application graph represents the minimal information that can be shared among the different managers to implement multi-concern autonomic management.

Consider a typical example, involving autonomic management of performance, security and power saving options in an application such as that of Fig. 4. A typical decision taken by the $\mathcal{AM}_P$ consists in varying the number of workers in the farm representing the second stage of the pipeline. For example, the number of workers can be increased to increase the throughput of the second stage and thus guarantee the user supplied performance contract. In this case, the decision of the $\mathcal{AM}_P$ will eventually lead to a different application graph. The new worker allocated will be labelled with some metadata representing, among other information, the resource where it will be mapped or the set of resources where the actual resource to host the worker will be taken from. The agreement with the other managers must be obtained in this case before committing the decision. $\mathcal{AM}_W$ may provide some priorities among the potential target resources for allocation of the new worker, in such a way that low consumption options are preferred. On the other hand, $\mathcal{AM}_S$ (the autonomic manager taking care of security concerns) may provide a binary mapping of the resources distinguishing those that are secure (i.e. those that can be reached using only private and trusted network segments) from those that are not. Eventually, $\mathcal{AM}_P$ may decide to allocate the new worker on a low consuming, secure resource (with no additional effort), on a low consuming, non-secure resource (with provision for encryption of communications) or on a high consuming, non-secure resource (again, providing for encryption). In all cases, the common level of agreement with the other managers will be on the final application graph. Even where no consensus can be reached among the different managers (e.g. no secure resources found, user contract asking for completely secure computations, impossibility to use alternative secure protocols) the eventual agreement will be on retaining the original graph, thus representing the fact that the decision by $\mathcal{AM}_P$ has been aborted.

## 3.3 Impact of local decisions on global application management

Having stated that a consensus has to be reached on the resulting application graph (with metadata) *before* committing any decision, we now consider how such consensus may be built. In particular, we discuss how the consensus process can be established and implemented; and the possible results of the consensus process.

### 3.3.1 Consensus building

Consensus building must be implemented as a two-phase process. In the first phase, the autonomic manager whose control cycle has identified that a decision has to be implemented as a consequence of some triggering event (here we assume it will be $\mathcal{AM}_1$) must initiate the consensus building, either by interacting with $\mathcal{AM}_0$ (SM case) or with all the other managers ($\mathcal{AM}_2$ to $\mathcal{AM}_m$, CM case). In the second phase, $\mathcal{AM}_1$ should await for the consensus results and, depending on their nature, either commit the decision (i.e. execute the actions in the plan associated with the decision) or abort it.

The intent of the two-phase protocol for consensus building is clear: no decision may be taken locally if the management of other concerns may be affected by the results of the decision. This in turn has two consequences:

1. decisions can be assigned to one of two classes: *independent* decisions, i.e. those not affecting the behaviour of other autonomic managers handling different concerns, and *interfering* decisions, i.e. those (potentially) having an impact on contract maintenance by other concern managers. For example, a decision to change the implementation of a parallel activity already mapped on a given processing resource from single to multi-threaded will most likely be an *independent* decision. On the other hand, a decision to migrate an already mapped parallel activity or to start a new parallel activity will be *interfering* decisions. In this case, new processing resources have to be recruited and that will typically affect contract maintenance by managers concerned with security, power management, etc.

2. decisions taken by $\mathcal{AM}_1$ could have several alternative equivalent implementations (i.e. plans and sequences of actions implementing the decision at $\mathcal{AM}_i$) including

    - plain implementation of the decision, i.e. no modification is made with respect to the implementation plan prepared by $\mathcal{AM}_1$ as a consequence of the answers provided by the other managers, and
    - "adjusted implementation" of the decision, i.e. an implementation whose actions have been modified according to the requirements gathered from the $\mathcal{AM}_j$ ($j \neq i$) in order to ensure maintenance of the whole set of contracts provided to the different managers rather than taking into account only concern $\mathcal{C}_1$.

    Typically, *independent* decisions will lead to the execution of unmodified implementation plans, whereas *interfering* decisions will lead to *adjusted* implementations.

Clearly, the necessity to provide "adjusted" implementation plans at manager $\mathcal{AM}_i$ raises a compositionality issue: if $\mathcal{AM}_i$ only had to take care of concern $\mathcal{C}_i$, no adjustment would be needed to its implementation plans. Adjustments are only needed when other concerns ($\mathcal{C}_j$, $j \neq i$) are taken into account. It is therefore clear that adjustments will depend on the nature of the $\mathcal{C}_j$. Thus $\mathcal{AM}_i$ will be no longer independent of the other managers/concerns.

In order to solve this issue, we propose the following methodology:

- A decision $\mathcal{D}_j$ taken by a manager $\mathcal{AM}_i$ is implemented with an ordered list of actions $a_{j1}, \ldots, a_{jk_j}$. This ordered list of actions is the implementation *plan* of decision $\mathcal{D}_j$.

- The granularity of the actions is the finest possible preserving the independence of each of the actions themselves.

- Actions are labelled as *independent* or *interfering* in the sense described above[6].

- Taking into account the overall set of concerns $\mathcal{C}_j$, $j \neq i$ considered in addition to $\mathcal{C}_i$, for each *interfering* action $a_k$ [7] one or more substitute plans $a_{k1}, \ldots, a_{ki_k}$ are prepared that have the same effect as $a_k$ with respect to the concern $\mathcal{C}_i$ but also accomplish some property required by other managers $\mathcal{AM}_j$, $j \neq i$.

---

[6] a decision is *interfering* if and only if it is implemented with a plan including at least one interfering action

[7] or for each sequence of actions $a_{k-m}, \ldots, a_{k+n}$ containing at least one interfering action $a_k$

- Finally, the consensus building phase will be modified as follows: the managers informed of decision $\mathcal{D}_k$ by manager $\mathcal{AM}_i$ will eventually report back to $\mathcal{AM}_i$ either an $ACK$ message or a $needProperty(propName_j)$ message, where $propName_j$ is one of the "other concern" properties $\mathcal{AM}_i$ is able to deal with[8]. If no suitable $propName$ is available at $\mathcal{AM}_i$ to deal with what is required by the other manager, a $NACK$ message will be returned that will serve to block the execution of $\mathcal{D}_k$ by $\mathcal{AM}_i$.

### 3.3.2 Consensus results

The consensus building process outlined in the previous section has already included mention of the various outcomes that can result from this phase.

The smoothest outcome is the one where $\mathcal{AM}_i$, seeking consensus on decision $\mathcal{D}_k$, gets from other managers (CM case) or from $\mathcal{AM}_0$ (SM case) only $ACK$ messages. This will be the result both in the case of an *independent* $\mathcal{D}_k$, and of an *interfering* $\mathcal{D}_k$ which at the moment does not cause any conflict with the policies implemented by the other managers.

The second case is in a sense the opposite of the first: $\mathcal{AM}_i$ gets at least one $NACK$ message back from one of the other managers. In this case the decision $\mathcal{D}_k$ will be *aborted* and manager $\mathcal{AM}_i$ must attempt to determine some other strategy (if any) to address the situation that triggered decision $\mathcal{D}_k$.

The last, and most interesting (and challenging) case, is that where $\mathcal{AM}_i$ gets all $ACK$ or $needProperty$ messages back from the other managers. Here we should distinguish two further sub-cases:

- There is a single $needProperty(propName_i)$ message. In this case, $\mathcal{AM}_i$ should simply implement the substitute plans for the interfering actions in the original $\mathcal{D}_k$ plan corresponding to $propName_i$[9].

- There are multiple $needProperty$ messages from the other managers. In this case $\mathcal{AM}_i$ should first determine which substitute implementation plans should be used and then consider whether the simultaneous usage of all of these substitute plans is still consistent. If it is consistent, the resulting new implementation plan will be executed. If not, $\mathcal{D}_k$ will be aborted.

Once the final plan implementing $\mathcal{D}_k$ has been determined (consensus having been achieved), the execution of the plan (i.e. the execution of the sequence of actions $a_1, \ldots, a_n$ in the plan) involves a modification of the application graph (the structure of the graph and/or the associated metadata). This modification

---

[8] these extra properties can be investigated by any $\mathcal{AM}_{i'}$ ($i' \neq i$) using introspection facilities on manager $\mathcal{AM}_i$

[9] these alternative plans should exist, otherwise $\mathcal{AM}_i$ should not have exposed $propName_i$ through its introspection facilities and therefore the involved manager should have sent back a $NACK$ message instead

has to be notified to all the other managers so that they can maintain a consistent view of the system. Moreover, the execution of the plan $a_1, \ldots, a_n$ has to be implemented as an *atomic* procedure. This means that any further decision taken by other managers should be processed only after finishing action $a_n$ and releasing the atomic action lock. In turn, all of this process obviously requires a distributed coordination mechanism. To avoid running a complicated and costly distributed coordination protocol, we can consider here to have the application graph controlled by $\mathcal{AM}_0$ in a SM implementation of the multiple concern management, and to have the single $\mathcal{AM}_i$ communicating the actions in the agreed plan to $\mathcal{AM}_0$ in such a way that these actions can be executed directly by $\mathcal{AM}_0$.

## 3.4 Initialization of the $\mathcal{AM}$ hierarchy

We assume that the user submits QoS contracts to the different $\mathcal{AM}$ s provided with the behavioural skeleton framework. These contracts describe the user's (non-functional) requirements that have to be guaranteed by the behavioural skeleton implementation of the user application.

We assume the user provides these contracts in such a way that:

- The order of the contracts establishes a priority among the managers. Thus, if the user provides contracts $QoS_1, \ldots, QoS_k$ (in order), only the managers dealing with concerns $\mathcal{C}_1$ to $\mathcal{C}_k$ will be activated and the decisions of manager $i$ will have precedence over the decisions of manager $i+h$. The relative ordering among managers and, consequently, among manager decisions can be used to solve conflicts when multiple decisions are communicated for consensus or even to impose an ordering on the substitute plan implementations when multiple *needProperty* messages have been directed to the $\mathcal{AM}_i$ seeking consensus on $\mathcal{D}_j$.

- The first contract $QoS_1$ determines which manager is in charge of establishing the initial application implementation configuration. This is particularly important as multiple concern management needs a starting configuration to initiate the autonomic management activities. Consider the case where performance, security and power saving concerns are of relevance. The same application will be configured to use the maximum number of powerful nodes, if run under the sole control of $\mathcal{AM}_P$, on a number of secure nodes, if run under the control of $\mathcal{AM}_S$, or on a number of low consumption nodes, if run under the control of $\mathcal{AM}_W$. In the three cases, the number of processing elements used may vary as well as the overall performance of the application.

## 3.5 Rule-based multi-concern autonomic manager implementation

In earlier work we demonstrated the suitability of business rule management frameworks for implementing autonomic managers handling a single concern [2]. A business rule framework implements a system of

$$P(x_1, \ldots, x_n) \to a_1; \ldots; a_k$$

(pre-condition ($P$) action) rules. When executed, the precondition part of all the rules is evaluated. Those rules that have a precondition holding true are fired[10]; that is, the corresponding action part is executed.

In particular, the classical control loop implemented by each manager may be implemented in such a way that:

- the monitor phase is implemented by gathering the current values of the variables used in the pre-condition parts of the rules;

- the analyse and plan phases correspond to evaluating which pre-conditions are satisfied and choosing one of the corresponding rules, possibly using some priority-based ordering;

- the execute phase is implemented by simply executing the action set (the implementation plan) in the right hand side of the rule identified in the previous step.

This was shown to work well when a single manager is considered. Now the idea can be adapted to the multi-concern management as follows:

- each rule originally present in the rule set implemented by $\mathcal{AM}_i$ *in isolation* is transformed into two distinct (classes of) rules:

  - a rule with the same pre-condition hosting as action part the consensus building start-up actions;

  - one or more rules with a pre-condition evaluating the responses provided by the other managers in the consensus building phase, and as the action part the original implementation plan or one of the *adjusted* plans.

- specific rules are added to deal with $NACK$ answers. These rules may include priority reordering within the manager rules, as well as new rules exploiting the available accumulated knowledge to deal with the new situation[11].

---

[10] possibly using some ordering based on priorities
[11] we assume here that some "learning" technique is used

# 4 Sample case study

We consider here a case study, to illustrate the concepts and the methodology discussed above.

Consider the application whose schema is depicted in Fig. 1, and assume two distinct non-functional concerns are handled by two autonomic manager hierarchies associated with the BS used: security and performance tuning.

Let us assume that the QoS contracts provided by the user are:

1. `secureData()`, directed to $\mathcal{AM}_S$ and specifying that all the data transfers involving remote nodes must be secured, and

2. `minThroughput(1 task/sec)`, directed to $\mathcal{AM}_P$ and specifying that the parallel application is expected to deliver at least one result per second.

As the first contract is directed to $\mathcal{AM}_S$, the autonomic manager dealing with security will handle the initial configuration of the program, i.e. it will define the initial application graph. Not being concerned with performance, $\mathcal{AM}_S$ will set up a graph using the default values for all those parameters that have not been specified by the user. In this case, the parallelism degree of the task farm will be set to some default value (say 4) and there will be no grouping of pipeline stages. Thus, the application graph will be a graph $\mathcal{G} = \langle N, A \rangle$ with:

$$N = \{n_{s1}, n_e, n_{w_1}, \ldots, n_{w_4}, n_c, n_{s3}\}$$
$$A = \{(n_{s1}, n_e), (n_e, n_{w_1}), \ldots, (n_e, n_{w_4}), (n_{w_1}, n_c), \ldots, (n_{w_4}, n_c), (n_c, n_{s3})\}$$

$\mathcal{AM}_S$ will try to select nodes $n_i$ that belong to trusted domains (i.e. domains that can be reached through trusted interconnections and hosting trusted nodes). If this is not possible, nodes from untrusted domains will be selected and metadata will be inserted in the application graph to state that the arcs leading to the untrusted nodes should be secured.

Once the initial application graph has been produced by $\mathcal{AM}_S$, it will be mapped onto the target architecture and the application will be started. After application start, metadata will be added to the application graph modelling node placement (e.g. *location($n_i$, ip_address$_j$)*), resource characterization (e.g. *nodeProp($n_i$, opSys(Linux), procType(dualcore), ...)*), etc. This metadata will be used to derive variables and values used in the pre-conditions as well as in the action part of the manager rules. Metadata also represent *de facto* the actual mapping of the abstract application graph to real resources.

Both $\mathcal{AM}_S$ and $\mathcal{AM}_P$ will start their control loops. $\mathcal{AM}_S$, being solely responsible for the initial allocation, will have no rules triggered and therefore will not execute any action affecting the system. On the other hand, $\mathcal{AM}_P$ will immediately evaluate the performance achieved by the program and this, in turn, will make some rules fireable if the performance is not the expected one[12]. Sample

---
[12]according to the user-supplied QoS contract

rules used in a hypothetical stand-alone $\mathcal{AM}_P$ should include the following rules for farms:

| Name | Rule |
|---|---|
| $Farm_{inc}$ | $priority(x)$, <br> $instanceof(farm)$ & $T_{arr} > QoS$ & $Throughput < QoS$ <br> $\rightarrow findNewResource, allocateNewWorker,$ <br> $\quad connectNewWorker$ |
| $Farm_{dec}$ | $priority(x)$, <br> $instanceof(farm)$ & $Throughput \gg QoS$ <br> $\rightarrow removeWorker$ |

($priority(x)$ denoting the fact that the rule has priority $x$, $T_{arr}$ being the inter-arrival time of tasks to the farm and QoS being the throughput contract issued by the user). These two rules will be different in an $\mathcal{AM}_P$ that is aware of the fact that it is managing performance while some other manager ($\mathcal{AM}_S$) is managing another concern. In this case, according to what we stated in Sec. 3 they should be of the form:

| Name | Rule |
|---|---|
| $Farm_{inc^{PH1}}$ | $priority(x)$, <br> $instanceof(farm)$ & $T_{arr} > QoS$ & $Throughput < QoS$ <br> $\rightarrow findNewResource, askConsensus(G', R')$ |
| $Farm_{inc^{PH2}}$ | $priority(x)$, <br> $ackFromAll \rightarrow allocateNewWorker, connectWorker$ |
| $Farm_{inc^{PH2}}$ | $priority(x)$, <br> $ackFromAll \& needProperty(security)$ <br> $\rightarrow allocateNewWorker, connectSSLWorker$ |
| $Farm_{inc^{PH2}}$ | $priority(x)$, <br> $nackConsensus \rightarrow lowerPriority(Farm_{inc})$ |
| $Farm_{dec}$ | $priority(x)$, <br> $instanceof(farm)$ & $Throughput \gg QoS$ <br> $\rightarrow removeWorker$ |

(where $G'$ is the new application graph resulting from the decision taken in the rule, $R'$ is the newly recruited resource).

In this case we assume the use of priorities to smooth the effect of aborted rules. Consider the example above. For the sake of simplicity, we omit other rules relating to autonomic management of performance in the task farm behavioural skeleton. However, it would be most likely the case that other rules exist that also happen to be fireable when rule $Farm_{inc^{PH1}}$ is fireable, i.e. when we have sufficient tasks to compute but still do not succeed in meeting the QoS contract. For example, a rule whose effect is to move a farm worker from a slow resource to a faster resource may exist, or a rule changing the kind of task-to-worker scheduling adopted in the farm to speed up computation. Now, if a rule has been selected and eventually aborted (as in $Farm_{inc^{PH2}}$ third item), by lowering the priority of the rule aborted we make fireable (at the next control loop iteration) an alternative rule firing on the same pre-condition but previously ignored due to its lower priority. This is a mechanism ensuring fairness in rule selection in the presence on NACKs during the consensus building phase.

In classifying actions as being *independent/interfering* (Sec. 3.3) we consider actions such as $findNewResource, askConsensus, allocateNewWorker$ as *independent* while actions such as $connectWorker$ are regarded as being *interfering*. In fact, the way we connect a worker (e.g. the way we implement the connections hosting communications between $n_e$ and $n_{w_{new}}$ and between $n_{w_{new}}$ and $n_c$ impacts the security (confidentiality and integrity) of the communicated data or code. Indeed, if $\mathcal{AM}_W$ (power management) is also included, the *allocateNewWorker* action must be considered *interfering*: the choice of a resource from those available will lead to a particular power consumption that in turn will eventually affect the power management concern managed by $\mathcal{AM}_W$. It is worth pointing out that *allocateNewWorker* actions could have been considered to be interfering actions when taking into account only the existence of $\mathcal{AM}_S$. However, the choice of a non-secure node in place of a secure one can be tolerated provided the actions and plans used by $\mathcal{AM}_P$ can be "adjusted" as outlined in Sec. 3.3. This is actually what happens in the rules above where the plan $findNewResource, allocateNewWorker, connectWorker$ is substituted (after consensus) with the plan $findNewResource, allocateNewWorker, connect-SSLWorker$.

In the general case, the decision to label an action as *interfering* depends on the set of concerns $\mathcal{C}_j$ ($i \neq j$) involved in addition to the concern $\mathcal{C}_i$ of the manager where the actions will be executed. Also, it is worth pointing out that metadata associated with the element of the application graph may influence handling of *interfering* actions. If the metadata associated with the application graph allows $\mathcal{AM}_S$ to conclude that the node added by $\mathcal{AM}_P$ is a secure node, no "adjustment" will be necessary to the *interfering* action $connectWorker$, for example.

On the other hand, $\mathcal{AM}_S$ should have rules such as:

| Name | Rule |
|---|---|
| $Node_{new}$ | $priority(y)$, $consensusAsked(G')$ & $diff(G_{current}, G') = N$ & $nonSecure(N)$ $\rightarrow answer(needProperty(security))$ |
| $Node_{new}$ | $priority(y)$, $consensusAsked(G')$ & $diff(G_{current}, G') = N$ & $secure(N)$ $\rightarrow answer(ACK)$ |

Within this minimal scenario, let us consider what will happen if $Farm_{inc^{PH1}}$

becomes fireable. This is a local decision taken by $\mathcal{AM}_P$. As a consequence, $\mathcal{AM}_S$ will be informed of the decision as it will receive a proposal for a new application graph with the extra node $n_{w_5}$ and the related new arcs $(n_e, n_{w_5}), (n_{w_5}, n_c)$. The node will have associated some metadata stating where it will be placed[13]. Two cases arise: either the metadata already includes the kind of node (secure, non-secure), or the $\mathcal{AM}_S$ will be able to determine such information from the metadata and from the other information it already has (or it can derive by monitoring). If $\mathcal{AM}_S$ identifies that the new node is secure, eventually an ACK will be delivered to $\mathcal{AM}_P$. Being the only other manager (hierarchy) in the system, this in turn will fire rule $Farm_{inc^{PH2}}$ (first item) in $\mathcal{AM}_P$ and this would commit the decision. If $\mathcal{AM}_S$ determines that the new node is non-secure, $\mathcal{AM}_S$ will issue a $needProperty(security)$ call and as a consequence the second item of rule $Farm_{inc^{PH2}}$ will be fired. This rule has an action part that represents an "adjusted plan" in that action $connectNewWorker$ has been replaced by $connectSSLworker$. This action will have the same effect as the former one: provide communication channels corresponding to arcs $(n_e, n_{w_5}), (n_{w_5}, n_c)$. However the connections will be set up using secure socket layer (SSL) and therefore the plan established by $\mathcal{AM}_P$ may be eventually actuated.

Despite the minimal "$\mathcal{AM}_P/\mathcal{AM}_S$" scenario outlined above, the system will eventually converge to the steady state, i.e. to the state where a suitable number of resources is used (to ensure the $\mathcal{AM}_P$ QoS contract) and the resources are either secure ones or they are connected using secure mechanisms. In a more realistic scenario, with a much more complex set of rules in the autonomic managers, convergence to the steady state may be accomplished with more effective policies, including, for example, the possibility to use a theoretical number of workers in a farm computed according to the task farm performance models *and* to the monitoring data gathered during the manager monitoring phase.

Security and performance are not the only non-functional concerns we investigated in our case study. Power management (and in particular power consumption optimization) is another concern we considered. We do not wish to enter a detailed description of $\mathcal{AM}_W$ rules and policies, here. But, in order to appreciate the flexibility of the proposed approach, we will describe an example of $\mathcal{AM}_P/\mathcal{AM}_S$ interaction. Assume we are again in the situation described before, where a task farm manager decides to increase the farm parallelism degree by adding a new worker to the worker string. Let us suppose that the new worker is used to support the additional load caused by a "hot spot" in the user application. Let us also assume the resource recruitment eventually returns a markedly *non-green* resource, i.e. a processing element consuming a considerable amount of power. It is mostly likely that the $\mathcal{AM}_W$ will conclude the consensus phase with a NACK and this will be handled with the techniques described above in the context of a NACK coming from $\mathcal{AM}_S$. However, in this case alternative

---

[13]generated by the query to the resource manager hidden in the $findNewResource$ action

strategies can be implemented. In fact, unlike the security concern, the power saving concern is not a binary issue. Therefore more sophisticated agreement and consensus algorithms can be used. For example, $\mathcal{AM}_W$ may judge feasible the allocation of the new worker on the power hungry resource provided a trade-off in power usage may be found elsewhere by $\mathcal{AM}_P$. Such a power saving may derive, for example, from the temporary suspension of some non-functional activity, such as monitoring application behaviour in order to be able to collate execution statistics, so that the amount of power saved may be used to fuel the hot spot.

As a final point, we must mention that in autonomic management of single non-functional concerns (see [3]) the concept of *violation* has been demonstrated useful. A violation is raised by a manager to its parent node manager when the contract is violated and no rules are fireable. In $\mathcal{AM}_P$ the situation arises, for example, when the task farm manager in our sample application (the one in Fig. 1) identifies that it is not satisfying the QoS contract due to the fact that too few task are arriving from the previous pipeline stage in a unit of time. In this case a violation is raised to the pipeline manager. The pipeline manager, in turn, can send a new QoS contract to the first stage manager, stating that a higher output rate is required. Here we do not consider the effect of violations. We are at an early stage in handling multiple concerns. The activity required to implement the two-phase decision (decision → consensus → actuation of the plan) is already time consuming. If we assume a manager $\mathcal{AM}_j$ ($j \neq i$) could raise exceptions and enter a possibly long process leading to some reconfiguration of other parts of the parallel application, we will in all likelihood have a reduction in the reactivity of the process implemented by manager $\mathcal{AM}_i$ and, more seriously, we must take great care to implement violations and the corresponding corrective actions at the parent node to avoid deadlocks.

## 5 Related work

The IBM blueprint paper on autonomic computing has already established, in a slightly different context, the need to orchestrate independent autonomic managers [11].

In [9] strategies to handle performance and power management issues by autonomic managers are discussed. However the approach is much more oriented to the generic combination of target functions relating to the two non-functional concerns considered, rather than to the constructive coordination of the actions planned by the two managers.

A framework that can be used to reason on multiple concerns was introduced in [12]. Based on the concepts of state and action (i.e., state transition) adopted from the field of artificial intelligence, this framework maps three types of agent-hood concepts (action, goal, utility-function) into autonomic computing policies. Action policies may produce and consume resources, which are used by a *resource* *arbiter* (i.e. a super manager) to harmonize conflicting concerns. The framework, however, does not provide any specific support to policy design and distributed management overlay.

A similar approach was followed in [6], which also exploits the same policies (action, goal, utility-function) defined on the (Cartesian product of) *state* and *configuration* space of the system. These policies are extended with *resource-definition* policies, which specify how the autonomic manager exposes the system to its environment; this makes it possible to dynamically extend manager knowledge with other resources/parameters, possibly coming from other managers, thus supporting management overlay.

The architecture and functionality of the latter two autonomic systems clearly resemble (and extend) those of intelligent multi-agent systems [14]. Our work substantially achieves similar features (see also [2]) in the context of adaptive parallel and distributed systems. As a distinguished feature, our work aims at defining tools and a methodology to ease and semi-automate the design and the generation of autonomic management and the distributed management overlay of a system, which is outside the scope of the related work. Our aim is focussed on the exploitation of structure by way of the co-design of functional and non-functional features [4] through behavioural skeletons, which may be regarded as autonomic management overlay factories.

# 6 Conclusions & future work

In this work we discussed a general methodology that can be used to support autonomic management of multiple non-functional concerns in a behavioural skeleton framework. The methodology is based on coordination of decisions taken by mostly independent autonomic managers (each taking care of a single non-functional concern) through a two-phase consensus protocol. We also discussed how the methodology can be applied to a typical parallel/distributed use case.

While protocols and policies may be established to coordinate the activities of different concern managers, the main challenge lies in not being overwhelmed by the sheer complexity of their interactions. To this end, we need to exploit to the full the fact that the structure of the underlying skeleton is *known* and use this knowledge in marshalling the activities of the overlaid autonomic management structure. In addition to this broad challenge, there remain a number of specific areas that require attention:

- We did not take into account autonomic manager hierarchy here. This means the rules outlined earlier appear to be executed by "the" autonomic manager, while they are actually fired by instances of the autonomic managers associated with the hierarchy of behavioural skeletons comprising the application. In particular, the handling of violations (raised child to par-ent) mentioned earlier introduces activity along an axis orthogonal to the multi-concern one. Coordination across the two axes remains an open issue.

- We did not distinguish in detail between use of the SM or the CM coordination model among managers. The details are hidden in actions *askConsensus*, *consensusAsked*, *answer*, *ackFromAll*, etc.

- No details are given relating to extent and format of the data exchanged among manager instances.

- No hint is given as to how managers can discover each other during application set up, in the case where the CM cooperation model is used.

- Some additional mechanisms are needed in the rules described above to make $\mathcal{AM}_i$ associate the correct pre-condition to the consensus results. In other words, we should label consensus items in such a way that replies to $askConsensus(X)$ do not fire rules whose pre-conditions are meant to capture answers to a different $askConsensus(Y)$ request.

In order to verify "in the field" the feasibility of the approach proposed in this work, we are currently finalizing a prototype that allows use of POJOs as processes (i.e. as the parallel pattern part of a behavioural skeleton) and the JBoss rule engine to associate some form of autonomic management to each of the simplified BS in the system (i.e. to implement the autonomic part of a behavioural skeleton). The resulting runtime system is completely based on RMI to run remote processes and assumes the machines have all the required ports open and a common (user) account is available to use the resources via ssh/scp. As a result, the runtime is quite straightforward and, as a consequence, easy to verify. This will allow us to experiment with complex cases and to validate the multi-concern autonomic management methodology proposed.